\documentclass[fleqn,usenatbib]{mnras}

\usepackage[T1]{fontenc}
\usepackage{ae,aecompl}
\usepackage{bm}
\usepackage{graphicx}    

\usepackage{amsmath}	
\usepackage{amssymb}	
\usepackage{txfonts}
\DeclareMathOperator{\sech}{sech}
\DeclareMathOperator{\gd}{gd}

\title[Analytic Solution for Kozai-Lidov Evolution]{An Analytic Solution to the Kozai-Lidov Evolution Equations  }

\author[Lubow]{Stephen H. Lubow$^1$\thanks{Email: lubow@stsci.edu}
\\ $^{1}$Space Telescope Science Institute, 3700 San Martin Drive, Baltimore, MD 21218, USA\\}

\date{Accepted XXX. Received YYY; in original form ZZZ}

\pubyear{2021}

\begin{document}
\label{firstpage}
\pagerange{\pageref{firstpage}--\pageref{lastpage}} 
\maketitle

\begin{abstract}
A test particle in a noncoplanar orbit about a member of a binary system can undergo Kozai-Lidov oscillations in which tilt and eccentricity are exchanged. An initially circular highly inclined particle orbit can reach high eccentricity. We consider the nonlinear secular evolution equations previously obtained in the quadrupole approximation. For the important case that the initial eccentricity of the particle orbit is zero, we derive an analytic solution for the particle orbital elements as a function of time that is exact within the quadrupole approximation. 
The solution involves only simple trigonometric and hyperbolic functions. It simplifies in the case that the initial particle orbit is close to being perpendicular to the binary orbital plane. 
The solution also provides an accurate  description of particle orbits with nonzero but sufficiently small initial eccentricity. 
It is accurate over a range of initial eccentricity that  broadens at higher initial inclinations. In the case of an initial inclination of $\pi/3$, an error  of 1\%  at maximum eccentricity
occurs for initial eccentricities of  about 0.1.
\end{abstract}


\section{Introduction}
The evolution of a test (massless) particle orbit around a member of a binary system has
been the subject of many studies. Particle orbits that are initially sufficiently inclined
with respect to the orbit of the binary can undergo oscillations in which 
inclination and eccentricity are exchanged 
(\citealt{Kozai1962}; \citealt{Lidov1962}; see review by \citealt{Shevchenko2017}). An initially highly
inclined circular test particle orbit around a member of a  binary can periodically gain high eccentricity
when the inclination decreases. 
 The theory of such orbits has been extended in many ways
to include effects such as higher order (octupole) corrections to the gravitational effects of the binary \citep{Naoz2013,
Naoz2016},
tidal friction within stars \citep{Kiseleva1998, Fabrycky2007}, general relativity \citep{Eggleton2001, Fabrycky2007}, and particle mass \citep{Hamers2021}.

The dynamical effects of such Kozai-Lidov (KL) oscillations have found application to many areas of astronomy,
including asteroids \citep{Kozai1962}, artificial satellites \citep{Lidov1962, Tremaine2014},  the eccentricities of planets \citep{Takeda2005, Saleh2009, Dawson2014}, the production of close binaries and close orbiting (hot Jupiter) planets \citep{Mazeh1979, Wu2003, Anderson2016, Petrovich2016}, merging black holes \citep{Blaes2002, Miller2002, Safarzadeh2020}, blue straggler stars \citep{Perets2009, Antonini2016, Fragione2019}, and the evolution of inclined protostellar disks in binaries \citep{Martinetal2014b, Zanazzi2017, Lubow2017}.

The analytical studies of KL oscillations are typically carried out by developing secular orbital evolution equations
that average over the binary and particle orbital periods.  The perturbing
companion is taken to be far from the test particle so that its effects can be treated in the quadrupole or octupole approximation. By making this approximation, the orbit averaging for secular evolution can be accomplished analytically.
The secular equations can be analyzed by analytic methods to provide insight into the nature of these oscillations.
The analytic studies have emphasized the evolution properties
starting with general initial conditions.  \cite{Kinoshita2007} provides analytic solutions  in terms of elliptic integrals
and an infinite Fourier series expansion for the longitude of the ascending node. \cite{Fabrycky2007}  determine the accessible states of systems by the  taking advantage  of the two constants of motion.

In this paper, we concentrate on the important case of an initially circular inclined particle orbit within the quadrupole approximation using
the secular equations given in \cite{Kiseleva1998}. Section \ref{sec:eq} reviews these equations.
Section \ref{sec:sol} derives the analytic solution
that is expressed in terms of the simple trigonometric and hyperbolic functions. 
Section \ref{sec:hii} describes the behavior of the solution at high initial inclination. 
Section \ref{sec:loi} discusses how the analytic solution extends to low initial inclination. 
Section \ref{sec:num}  discusses  a comparison of the analytic solution with a numerical  solution.
Section \ref{sec:res} describes some phase portraits.
Section \ref{sec:conv} discusses the errors made in using the analytic solution in cases where the initial particle orbit
is noncircular.
Section \ref{sec:sum} contains the summary.


\section{Orbital Evolution Equations}
\label{sec:eq}

Consider a test (massless) particle that orbits a member of a binary with mass $M_1$ and is perturbed by an exterior companion of mass $M_2$.
The particle orbit is inclined relative to the binary orbital plane.
We adopt the quadrupole approximation for the secular evolution of the  particle orbital elements. The secular evolution equations for particle eccentricity $e$, inclination relative to the binary orbital plane $i$,
argument of periapsis $\omega$, and longitude of the ascending node $\Omega$, following equations 2 -5 in  \cite{Kiseleva1998}, are given by
\begin{eqnarray}
 \sqrt{1-e^2} \frac{d e}{dt'} &=& 5 e (1-e^2) \sin^2{i} \sin{\omega}  \cos{\omega} ,\label{e}\\
 \sqrt{1-e^2} \frac{d i}{dt'} &=& - 5 e^2  \sin{i} \cos{i} \sin{\omega}  \cos{\omega}, \label{didt}\\
  \sqrt{1-e^2} \frac{d \omega}{dt'} &=&  ( 5 \sin^2{\omega} -1) (e^2- \sin^2{i}) \nonumber   \\ && 
                                +1  -e^2 +\cos^2{i},  \label{domegadt}  \\
 \sqrt{1-e^2}  \frac{d  \Omega}{dt'} &=& -[1 + e^2 (5 \sin^2{\omega} -1)]\cos{i}, \label{Omega}
\end{eqnarray}
where $t'$ is a dimensionless time with $t' = t/\tau$ for time $t$, timescale
\begin{equation}
\tau = \frac{2 P_{\rm b}^2 (M_1+M_2)}{3 \pi P M_2} (1-e_{\rm b}^2)^{3/2},
\end{equation}
with binary orbital period $P_{\rm b}$, binary eccentricity $e_{\rm b}$,
and  particle orbital period  $P$. 

In addition, from  equations 6 and 7 in  \cite{Kiseleva1998}, there are two constants of motion
that are
\begin{eqnarray}
 (1-e^2) \cos^2{i} &=& {\rm constant},  \label{igen}\\
\left [5 e^2 \sin^2{\omega} +2 (1-e^2) \right] \sin^2{i}  &=& {\rm constant} . \label{omegagen} 
\end{eqnarray}

We take time $t'  =  0$ to be when
the eccentricity is maximum.  
Under the quadrupole approximation applied in this paper, the oscillation period is formally infinite with a weak logarithmic singularity
in the limit of small initial eccentricity $e_{\rm init}$.
This singularity is removed in octupole order, as discussed in  \cite{Kiseleva1998}.
The "initial" time $t'_{\rm init}$ is then formally at $t'=-\infty$. 
 The particle orbit is taken to be initially circular, $e_{\rm init}=0$ in the sense of a limit at $t' \rightarrow -\infty$.
We are interested in a solution for $e(t')$ for which  $e(t')>0$ at any finite time.  
The eccentricity is then small but nonzero
at any early finite time. There is also a solution for which  $e_{\rm init}=0$ at finite initial time. This point is discussed further in Section \ref{sec:loi}.

 For this case of an initially circular particle orbit these constants reduce to
\begin{eqnarray}
 (1-e^2) \cos^2{i} &=& \cos^2{i_{\rm init}},  \label{i}\\
  \left[5 e^2 \sin^2{\omega} +2 (1-e^2)  \right] \sin^2{i}  &=& 2 \sin^2{i_{\rm init}} . \label{omega} 
\end{eqnarray}

\section{General Solution for Initially Circular Orbit}
\label{sec:sol}

Applying Equations (\ref{i}) and (\ref{omega}) in Equation (\ref{e}), we obtain a solution for $e(t')$ subject to the initial
condition described in the previous section that can
then be used to determine the other orbital elements as a function of dimensionless time $t'$. The result is
that
\begin{eqnarray}
e\left(t' \right) &= & \frac{ q}{\sqrt{6}}  \sech{\left( q  \, t'\right)},  \label{es}\\
\cos^2({i(t')})  &=&  \frac{ 6 \cos^2{(i_{\rm init}) }}{6 - q^2 \sech^2{( q \, t')}}, \label{is} \\
\sin^2({\omega(t')})  &=&   \frac{4[ 3(1+ \cosh{(2 q \, t')})- q^2] }{12-7 q^2 + 3(4+q^2)\cosh{(2 q \, t')} }, \label{omegas} \\
\Omega{(t')} &=& \Omega_0 - \cos{(i_{\rm init})}  t' - \arctan{\left(\frac{q \tanh{(q \, t')}}{2 \cos{(i_{\rm init})} } \right)} \label{Omegas},
\end{eqnarray}
where $\Omega_0$ is the longitude of the ascending node at time $t'=0$ and
\begin{equation}
 q = \sqrt{1- 5 \cos{(2 i_{\rm init})}}. \label{q}\\
\end{equation}

Equation (\ref{Omegas})  for $\Omega$ contains a term that corresponds to a uniform 
precession that is present even at low inclinations (as is discussed in Section \ref{sec:loi}), as well as a nonuniform precession term that is 
due to the KL effect.  As discussed in Appendix \ref{sec:mercator}, this nonuniform precession term is closely related to the Gudermannian function and the Mercator projection (see \url{https://en.wikipedia.org/wiki/Gudermannian_function}).  

The KL oscillations take place for real values of parameter $q$ defined in Equation (\ref{q}).  Oscillations occur when
$0  < q \le \sqrt{6}$. This constraint is satisfied over a range of initial tilt angles for which $\cos(2  i_{\rm init}) <  1/5$ that corresponds to $141^\circ \la  i_{\rm init}   \la 39^\circ$ 
 \citep{Kozai1962, Lidov1962}.  
  Equation (\ref{is}) can also be written as
 \begin{equation}
 \cos^2({i(t')})  =  \frac{ 3( 6 - q^2)}{5 (6 - q^2 \sech^2{( q \, t')})}, \label{isa}
 \end{equation}
 which shows that $  \cos^2({i(t')}) \le 3/5$ for this range of $q$ values.  Equation (\ref{isa}) implies that  the inclination reaches a minimum value at $t'=0$  that is equal to the Kozai-Lidov critical angle, $\cos^2({i(0})) = 3/5$ or equivalently $\cos({2 \, i(0})) = 1/5$, as is well known.

 For any orbit in the KL regime, notice that by Equation (\ref{e}), the eccentricity $e$ is maximum in time ($de/dt'=0$)
 for
  \begin{equation}
  \omega(0) = \pm \frac{\pi}{2} \label{om0}.
  \end{equation}  
 This value of $\omega(0)$ agrees with the solution given by Equation (\ref{omegas}) which implies
 $\sin^2({\omega{(0)})}=1$. 
    Which of the two values of $\omega(0)$  is  realized depends on initial conditions.
  (With  $\omega = 0$, or $\pi$,  derivative $de/dt=0$ can also occur in Equation (\ref{e}),  but corresponds to an eccentricity minimum because $d^2e/dt'^2>0$.  That follows from applying Equation (\ref{domegadt}) 
  to the time derivative of Equation (\ref{e}).)
 Notice that Equation (\ref{omegas}) for $\omega$ does not contain an arbitrary constant of integration, as occurs in Equation (\ref{Omegas}). 
 The phasing of $\omega(t)$ is determined at the time of eccentricity maximum ($t'=0$), as given by Equation (\ref{om0}).
   
 From  the time derivative of Equation (\ref{e}) and Equation (\ref{om0}), it follows that $d\omega/dt (0)>0$. We apply this condition in  Equation (\ref{omegas}) along with
 Equation (\ref{om0}) to determine $\omega(t')$. In addition, we  determine $i(t')$ from Equation (\ref{is}) to
 obtain
 \begin{eqnarray}
 i(t')  &=& \arccos{\left( 
\frac{ \sqrt{6} \cos{(i_{\rm init}) }}{\sqrt{6 - q^2 \sech^2{( q \, t')}}} 
\right)}, \label{is1} \\
\omega(t') &=&  \pm \pi H(\pm t') -  {\text{sgn}}(t') \times \nonumber \\ &&    \arcsin{ \left( \sqrt{
\frac{4[ 3(1+ \cosh{(2 q \, t')})- q^2] }{12-7 q^2 + 3(4+q^2)\cosh{(2 q \, t')} } 
} \right)},  \label{omegas1} 
\end{eqnarray}
where  $H$ is the Heaviside step function.
These  functions and their derivatives vary smoothly in time.

At early times ($t' \ll 0 $) we have that
 \begin{eqnarray}
 e(t') &=& \frac{\sqrt{6}\, q}{3} \exp{(q t')},  \label{einit} \\
i(t') &=& i_{\rm init},  \label{iinit} \\
 \omega(t') &=& -(1\mp1)\frac{\pi}{2} +  \nonumber\\ && \arcsin{ \left( \sqrt{\frac{4}{4+q^2} } \right)},  \label{ominit} \\
 \Omega{(t')} - \Omega_0 + \cos{(i_{\rm init})}  t'  &=&  \arctan{\left(\frac{q }{2 \cos{(i_{\rm init})}  } \right)}.
\end{eqnarray}

 The KL effect can behave as an instability in eccentricity. A sufficiently inclined orbit with small initial eccentricity undergoes 
exponential growth in time at nearly constant inclination.
Equation (\ref{einit}) shows that the eccentricity grows exponentially at early times with a growth rate
 $ d\ln{e}/dt = q /\tau$ that agrees with equation 43 of \cite{Tremaine2014} and equation 9 of \cite{Lubow2017}
 \footnote[1]{There is a typo in equation 9 of \cite{Lubow2017}, the factor $ \sqrt{5 - 3 \cos^2{i}}$ should be
   $\sqrt{3 - 5 \cos^2{i}}$ .} (see also \cite{Tremaine2009})
   The KL effect can also be understood as involving a resonance, as will be discussed in Section \ref{sec:res}.

  \section{Solution at High Initial Inclination}
  \label{sec:hii}

We consider cases where the initial particle orbit is both circular and highly inclined.
We define angle
\begin{equation}
\theta = \pi/2 - i_{\rm init}  \label{theta}
\end{equation}
and assume that $0< \theta \ll 1 $.
We consider multiple timescales defined as
\begin{equation}
t_j = t'/\theta^j
\end{equation}
for integer $j$.
 Equations (\ref{es}),   (\ref{is1}), (\ref{omegas1}), and (\ref{Omegas}) respectively approximately reduce to
\begin{eqnarray}
e &=& \left(1- \frac{5}{6} \theta^2 \right) \sech{ (\sqrt{6} t_0) }, \label{ehi}\\
i &=& \arccos{\left( \sqrt{\frac{1}{\frac{5}{3} + \frac{25}{9} \theta^2 +6 t_1^2}+\theta^2
}\right)}, \label{ihi}\\
\omega &=& \pm \pi H(\pm t_1)  - {\text{sgn}}(t_1) \arcsin{ \left ( \sqrt{\frac{5 + 18 t_1^2}{5 + 45 t_1^2}} \right)}, \\
\Omega &=& \Omega_0 -  t_{-1} - \arctan{\left( 3 t_1 \right)} . \label{Omegahi}
\end{eqnarray}
We have retained each contributing term  to at least lowest order in $\theta$ in order
to obtain a reasonably accurate solution at all times. 
We see that orbital elements $i$, $\omega$, and $\Omega$   
undergo large changes on a relatively short timescale $\theta \, \tau \ll \tau$
from the time of the eccentricity maximum. The eccentricity varies only on the longer timescale $\tau$.

\begin{figure}
\includegraphics[width=8 cm]{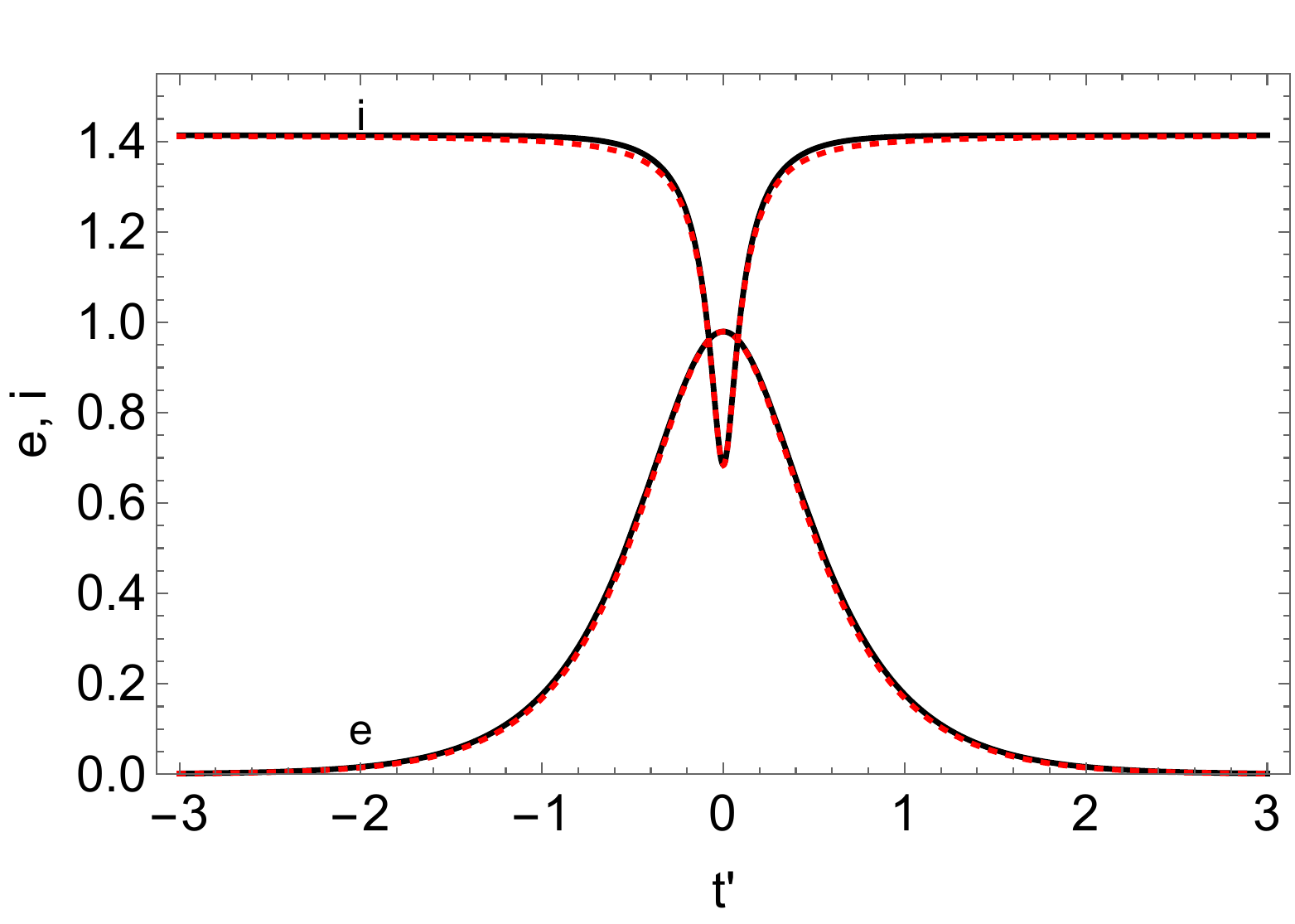}
 \hspace{-0.7cm}
 \includegraphics[width=8 cm]{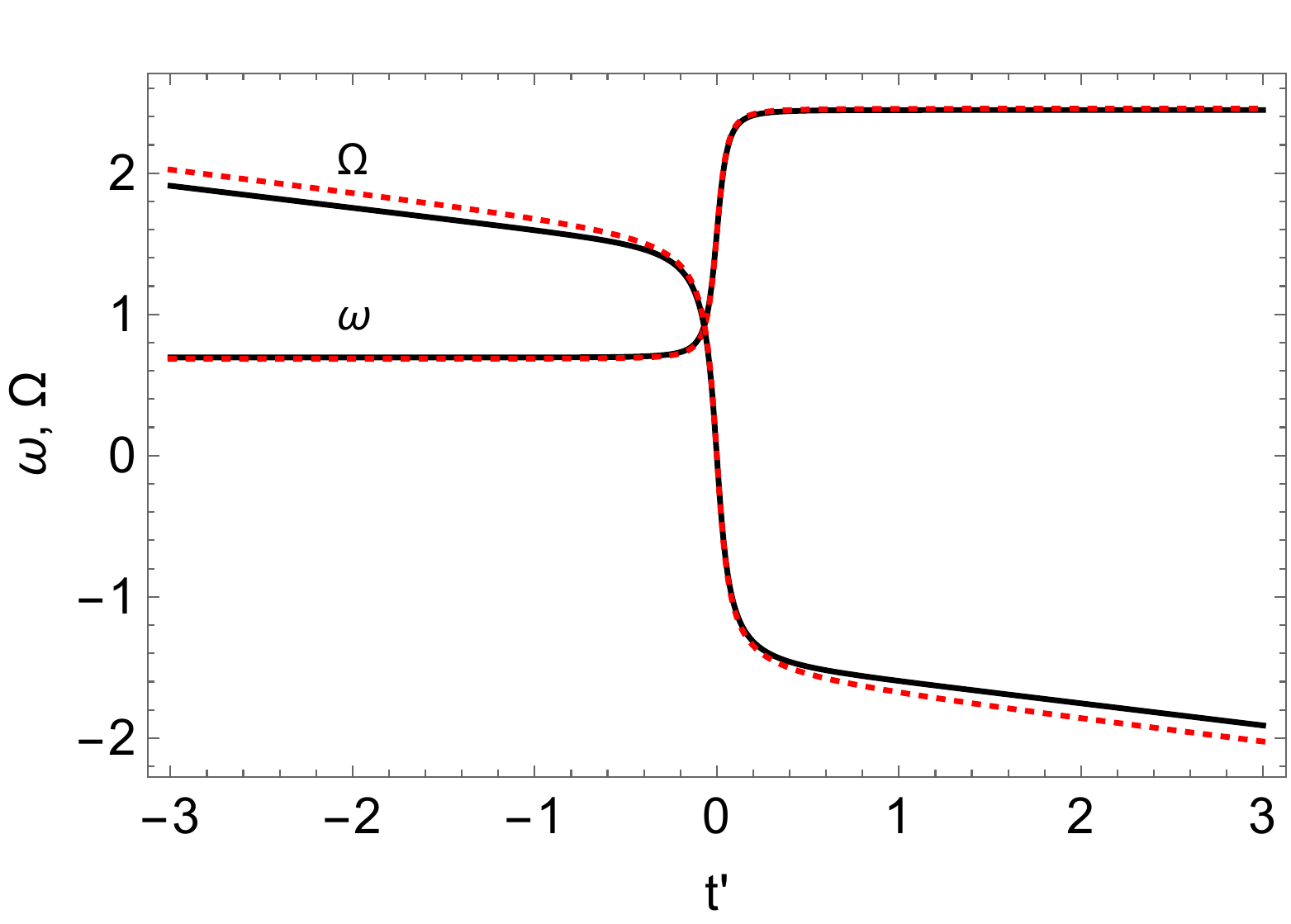}
\caption{Plot of orbital elements $e$, $i$, $\omega$, and $\Omega$ versus dimensionless time based on the full analytic solution in black solid lines given by Equations (\ref{es}), (\ref{is1}), (\ref{omegas1}), and (\ref{Omegas}), respectively,  and the approximate expressions in red dotted lines given by Equations (\ref{ehi}) - (\ref{Omegahi}). The initial inclination is $i_{\rm init}= 0.45 \pi$ and nodal phase constant $\Omega_0=0$.
}
\label{fig:ei}
\end{figure}

Figure~\ref{fig:ei} plots the orbital elements  with  $i_{\rm init} = 0.45 \pi$ and  $\Omega_0=0$  using
the full analytic expressions given in Section \ref{sec:sol} as black solid lines and the approximate expressions given above  by red  dotted lines. The approximate expressions are quite accurate with expansion parameter $\theta=0.05 \pi \simeq 0.16$ in this case. A shown in the figure, $i, \omega,$ and $\Omega$
undergo large changes within a small timescale, while $e$ does not vary as much on that timescale.

\section{Solution at Low Initial Inclination}
\label{sec:loi}

For any inclination, there is another solution to Equations (\ref{e}) - (\ref{Omega})
that begins with $e_{\rm init}=0$, where $t'_{\rm init}$ can be a finite or infinite initial time.
For this solution, we have
\begin{eqnarray}
e(t') &=& 0, \label{ezs}\\
i(t') &=& i_{\rm init}, \label{izs} \\
 \Omega{(t')} &=& \Omega_0 - \cos{(i_{\rm init})}  t'. \label{Omegazs}
\end{eqnarray}
Consideration of $\omega(t')$ is omitted for this solution, since the orbit
remains circular. For $i_{\rm init}$ within the KL angle range,
$q >0$ and the eccentricity given in Section \ref{sec:sol} grows exponentially
in time for arbitrarily small but nonzero initial values at finite time (see Equation (\ref{einit})). Therefore, this
constant zero eccentricity solution is unstable for $q>0$ and the solution in Section \ref{sec:sol}
then applies.

Both solutions coincide at $q=0$. That is, the solution in
Section \ref{sec:sol} and the above solution are the same
for $q=0$. Outside the KL angle range (including the case of low initial
inclination), Equations (\ref{ezs}) -
(\ref{Omegazs}) provide the solution. Notice that the solution
in Section \ref{sec:sol} is valid for all inclination angles if we take $q$
to be the real part of $q$. In that way, $q=0$ outside the 
KL angle range and Equations (\ref{es}), (\ref{is}), and (\ref{Omegas})
reduce to Equations (\ref{ezs}) - (\ref{Omegazs}).

\section{Numerical Verfification}
\label{sec:num}
The analytic solutions for the orbital elements $e$, $i$, $\omega$, and $\Omega$ given respectively by Equations (\ref{es}), (\ref{is1}), (\ref{omegas1}), and (\ref{Omegas}) were compared to numerical solutions.
The numerical calculations were carried out using Equations (\ref{e}) - (\ref{Omega}) together with initial
conditions provided by the analytic solutions at time $t'=-5$ and $i_{\rm init}=\pi/3$. The equations were integrated
from $t'=-5$ to $t'=5$.
 Both branches of $\omega$ given in  Equation (\ref{omegas1}) were tested.
  The calculations
 were carried out using NDSolve in Mathematica. 
 Over this time interval, the eccentricity $e$ ranges from about $1.3 \times 10^{-4}$ at the endpoints  to about 0.76 at the midpoint.
 The analytic and numerical results agreed to about $1\times10^{-6}$ for all orbital elements throughout this time interval.
 The errors are likely due to the limits of the precision in the numerical integration.

  \section{Phase Portraits}
\label{sec:res}

 \begin{figure}
\includegraphics[width=7 cm]{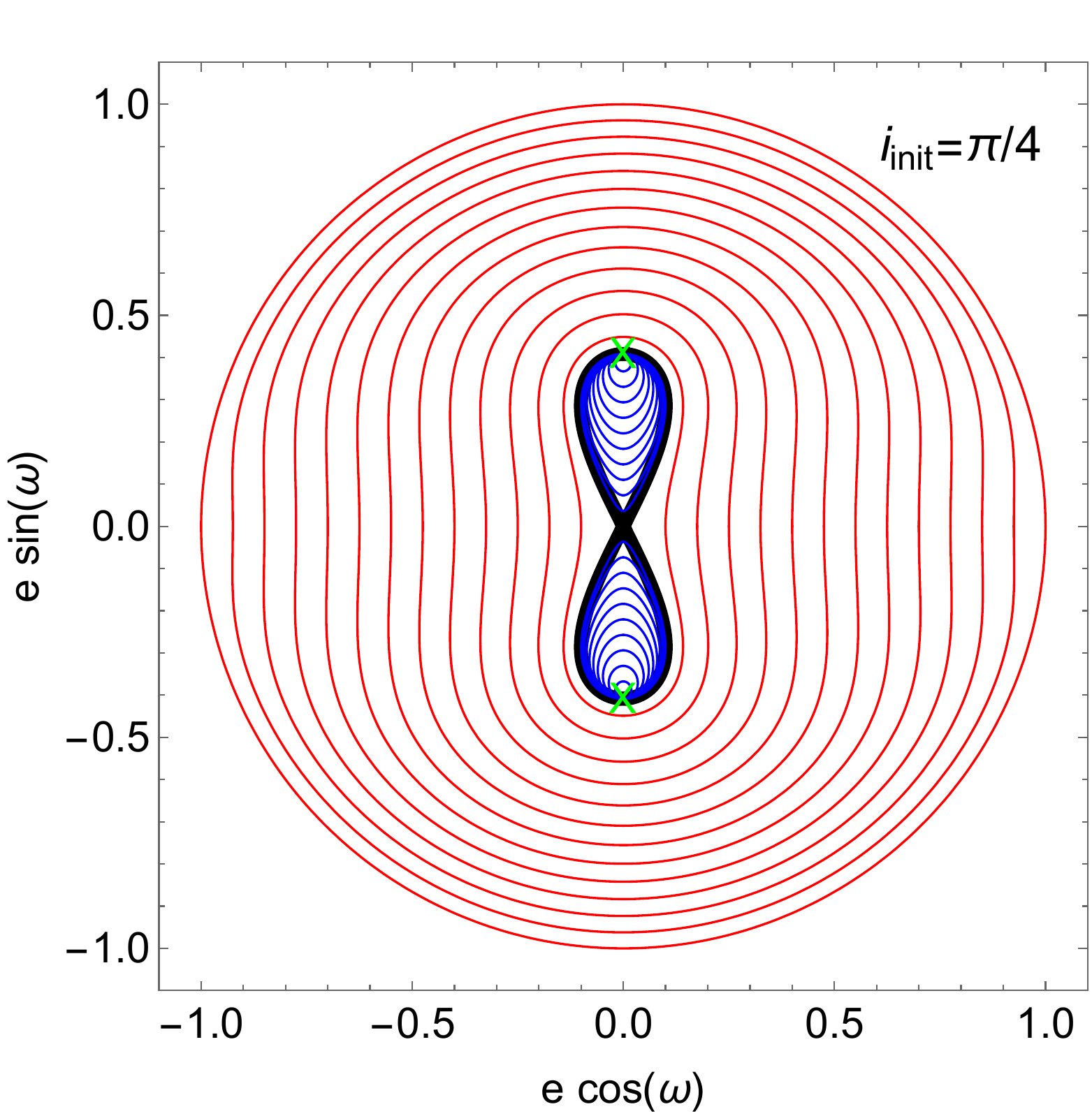},
\includegraphics[width=7 cm]{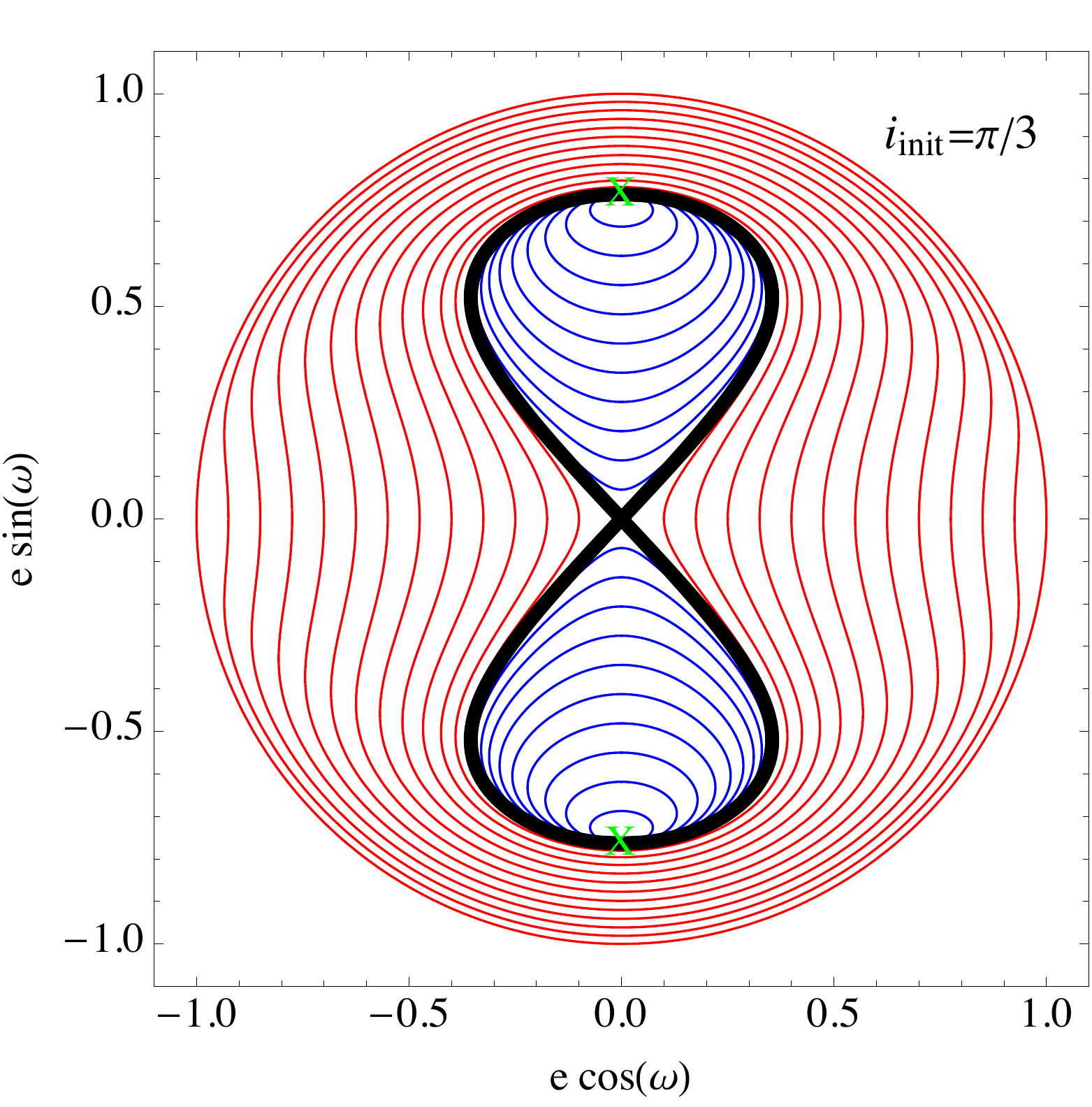},
\includegraphics[width=7 cm]{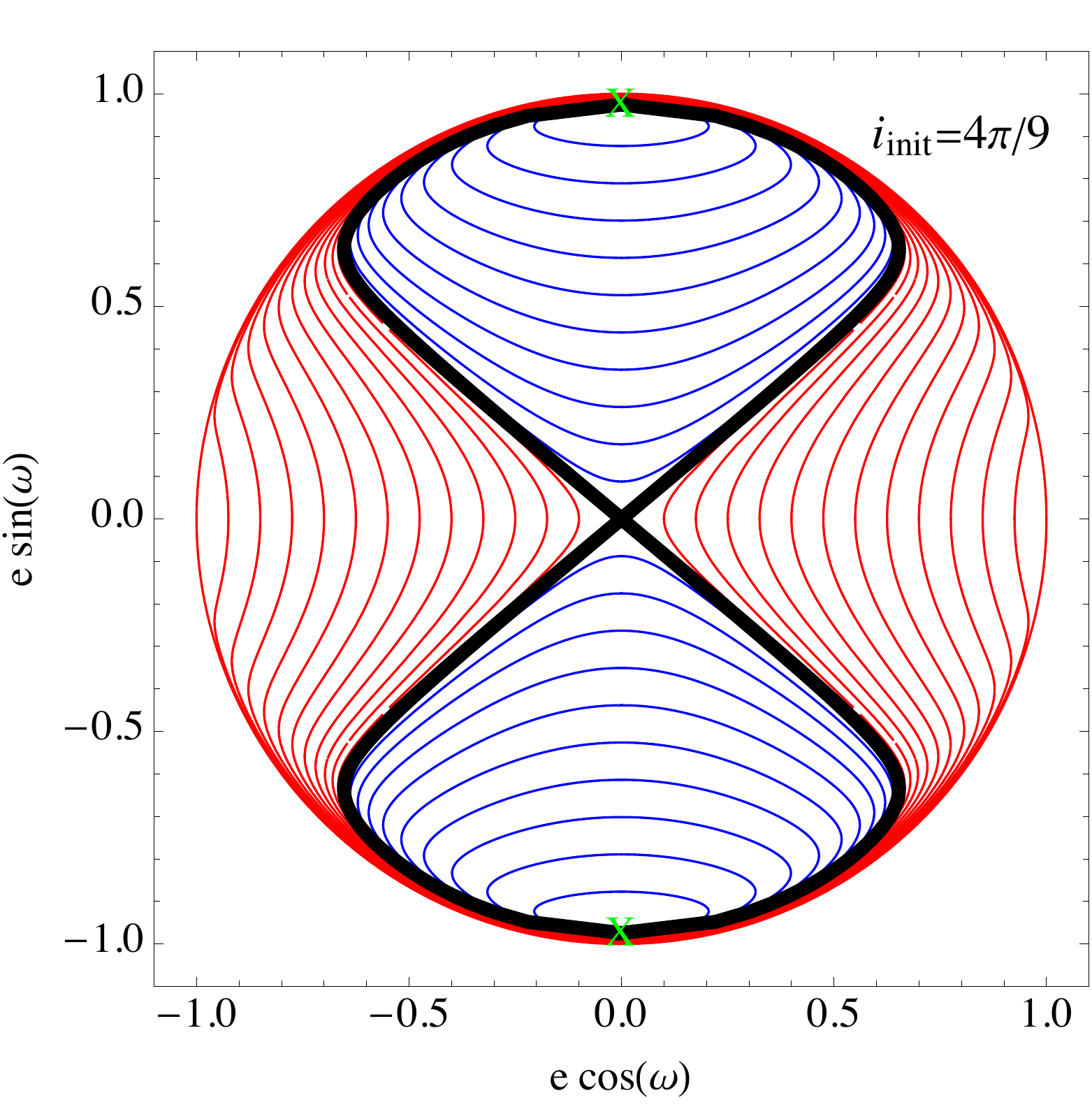}
\caption{ Phase portraits of  $e \cos{(\omega)}$ versus
$e \sin{(\omega)}$ for three values of the initial orbit inclination $i_{\rm init}$.
The heavy black line is a separatrix and corresponds to the solutions found in Section \ref{sec:sol}.
Orbits inside (outside) the separatrix all have $\omega_{\rm init} = \pm\pi/2$ (0). The different orbits have
different values of $e_{\rm init}$. The green X marks the maximum eccentricity on the separatrix.
}
\label{fig:phport}
\end{figure}

Figure~\ref{fig:phport}  plots some phase portraits of $e \cos{(\omega)}$ versus
$e \sin{(\omega)}$ that result from solving Equations (\ref{igen})
and (\ref{omegagen}).  There is a separate phase portrait for each of three values of $i_{\rm init}$.
For any point in a plot, the value of the eccentricity $e$ is its distance from the origin 
and the  argument of periapsis $\omega$ is its polar angle from the horizontal. The heavy black line in each plot
passes through
the origin and therefore corresponds to an orbit that has zero eccentricity at some time.   The time dependence
of this orbit is described by the analytic solution given in Section \ref{sec:sol}. 
The heavy line is also a separatrix in each plot.

We consider the initial values with subscript {\rm init} for each orbit to occur where the eccentricity is minimum (closest
point to the origin along the plotted orbit).
The plotted orbits inside the separatrix are chosen to have $\omega_{\rm init}=\pm \pi/2$. They are librating orbits, orbits that
undergo a limited range of $\omega$ less that $2 \pi$.  The orbits outside the separatrix
 have  $\omega_{\rm init}=0.$  They are circulating orbits, orbits that
undergo a full range of $\omega$ equal to $2 \pi$.
  The different plotted orbits have different values of $e_{\rm init}$.
The plotted orbits inside and outside the separatrix
have nonzero eccentricities at all times. 

The existence of the librating orbits reflects the resonant nature
of the KL oscillations \citep[e.g.,][]{Malhotra2012, Shevchenko2017}.  The libration is associated with an apsidal-nodal
resonance in which the average value of $d \omega/dt =  d \varpi/dt - d\Omega/dt$ is zero over the libration period,
where $\varpi$ is the  longitude of the periapsis.

The maximum eccentricity on the separatrix occurs at points marked by a green X.
 The orbits inside the separatrix, which all have $\omega_{\rm init}=\pm \pi/2$,
converge to this point with an ever decreasing range of $\omega$ values.

 \section{Convergence to the Analytic Solution}
\label{sec:conv}
We examine whether the analytic solutions are stably reached.
That is, whether numerical solutions with different initial conditions would approach the analytic
solutions in Section \ref{sec:sol} over time.  By doing so, we obtain 
an estimate of the accuracy of using the analytic solution given
by Section \ref{sec:sol} in cases where the initial eccentricity $e_{\rm init}$ is nonzero.

There are different ways to define the convergence to the analytic solution.
 Testing for convergence involves a set of an initial parameters that are varied and a measure
 of error at a later time. We consider initial values of $e_{\rm init}>0$, $i_{\rm init}$,
 and $\omega_{\rm init}$.

We  examine the numerical
solutions of Equations (\ref{e}) - (\ref{Omega}) and values of $e_{\rm init}$ and $\omega_{\rm init}$  that differ from the analytic solution
of Section \ref{sec:sol} for a given value of $i_{\rm init}$. Of interest, is whether those solutions
approach the analytic solution over time.
In carrying out the numerical integrations for this test, we do not know in advance the time of maximum eccentricity. Instead, we take the starting time of the numerical integrations as $t'=0$ and
determine the time of  peak eccentricity  $t'_{\rm peak}$.  
The comparison with the analytic
solutions can then be made  by a simple time shift to redefine time $t'_{\rm peak}$ of numerical solutions to time zero, $t'=0$. 
The analytic solution has an infinite period, while the numerical solutions have a finite period.
We compare results over the first full period of the numerical solutions centered at the time when the
eccentricity is maximum.

\begin{figure}
\includegraphics[width=8 cm]{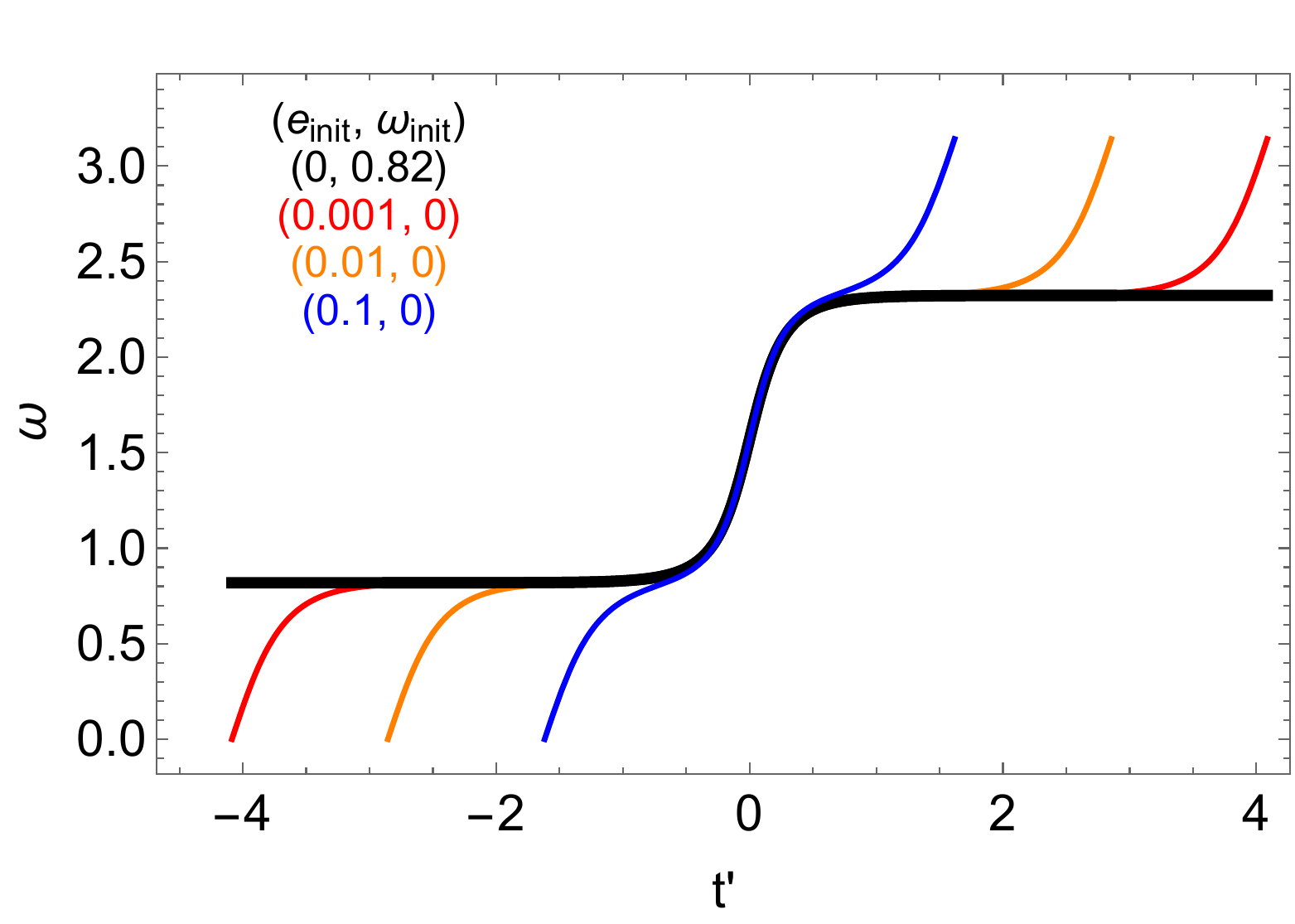}
\includegraphics[width=8 cm]{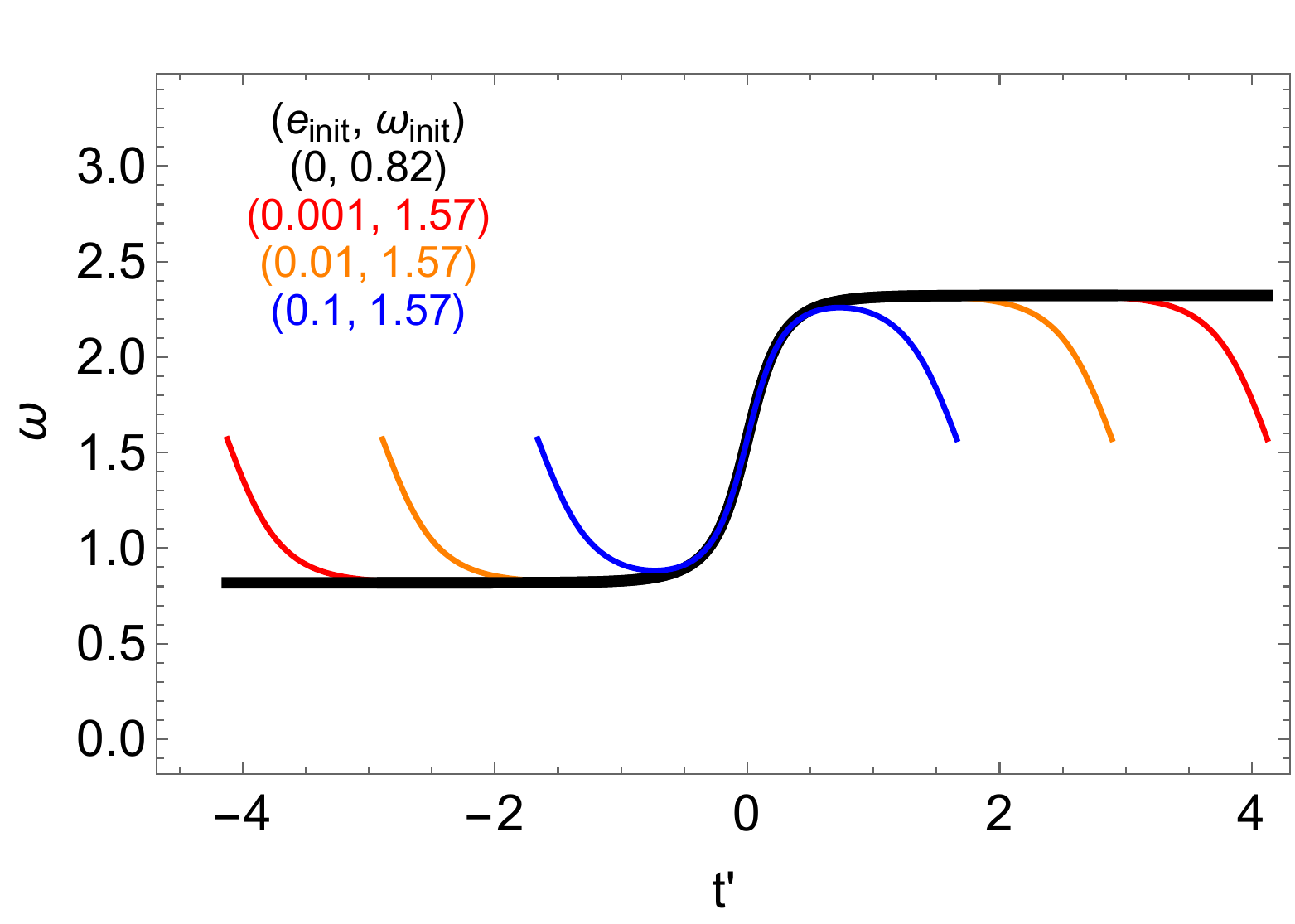}
\caption{ The plot shows the effect of changing the initial condition on the argument of periapsis $\omega$
from the value given by the analytic solution of $ \omega_{\rm init} \simeq 0.82$ 
to  $ \omega_{\rm init} =0$  (upper plot) and $ \omega_{\rm init} =\pi/2$ (lower plot).
The initial inclination is $\pi/3$ in all cases.
The black line shows the analytic solution given by Equation (\ref{omegas1}).
Nonblack colored lines are the numerical solutions for the argument of periapsis 
$\omega$ over the first oscillation in integrating Equations (\ref{e}) - (\ref{Omega}) as a function
of time relative to the time of the eccentricity maximum. The
initial values for eccentricity  $e_{\rm init}$ are 0.001 (red), 0.01 (orange), and 0,1 (blue).
 }
\label{fig:omt}
\end{figure}

\subsection{Convergence of $\omega(t')$}

We investigate the convergence of $\omega(t')$.
 We consider  cases
with starting values $i_{\rm init}=\pi/3, \omega_{\rm init}=0, \pi/2$, and $\Omega_0=0$. Three values
of initial eccentricity are applied that are $e_{\rm init} =0.001, 0.01,$ and $0.1$. 
Orbits with $\omega_{\rm init}=0$ are outside the separatrix as shown in Figure~\ref{fig:phport},
while orbits with $\omega_{\rm init}=\pi/2$ are inside the separatrix.
The analytic solution has the same
initial inclination, but with $e_{\rm init}=0$ and $\omega_{\rm init}=\arcsin{(2 \sqrt{2/15})} \simeq 0.82$ at $t'=-\infty$ as given by Equation (\ref{ominit}).
The results plotted in Figure~\ref{fig:omt}  show that the numerical results for $\omega(t')$ are close
to the analytic results midway through the oscillation, as expected because $\omega(0)=\pm \pi/2$ in all cases as seen
in Equation (\ref{om0}). 

 $\omega(t')$  reaches a value close to the analytic solution and breaks away at later times due to periodicity.
The time interval over which the numerical 
results agree well  with the analytic results increases with decreasing $e_{\rm init}$.
 Therefore, the numerical results for $\omega(t')$ converge
towards the analytic result as $e_{\rm init}$ approaches zero, for both values of $\omega_{\rm init}$.
However, the infinite time available as $e_{\rm init}$ approaches zero is an artifact of the quadrupole approximation.
In reality, the convergence would be limited in the case that the higher order gravitational moments 
are taken into account.

\subsection{Convergence of $e(0)$}
\label{sec:e0}

We apply a 
 measure of error that is the difference between the maximum eccentricity of an orbit with some set of initial
 conditions
 from the maximum eccentricity of the analytic solution of Section \ref{sec:sol} that has the same value of  $i_{\rm init}$.
 In particular, we want to know how rapidly this eccentricity difference
approaches zero as  $e_{\rm init}$ approaches zero. Recall that we define $t'=0$ to be the time
at which the maximum eccentricity occurs. We are then comparing solutions at the same time $t'=0$. We know from Equation (\ref{om0}) that this occurs at
$\omega = \pm \pi/2$, as seen in Figure~\ref{fig:phport}.

We denote the analytic solution for maximum eccentricity given by the analytic solution as 
\begin{equation}
e_{\rm amax}(i_{\rm init}) = \frac{\sqrt{1-5 \cos{(2 i_{\rm init})}}}{\sqrt{6}} \label{eamax}
\end{equation}
that follows from Equation (\ref{es}). 
The maximum eccentricity for an orbit with $e_{\rm init}$ nonzero
 is denoted by $e_{\rm nmax}(e_{\rm init}, i_{\rm init}, \omega_{\rm init})$.
We determine the error
\begin{equation}
\Delta e (e_{\rm init}, i_{\rm init}, \omega_{\rm init})  = |e_{\rm nmax}(e_{\rm init}, i_{\rm init}, \omega_{\rm init})-e_{\rm amax}(i_{\rm init})|, \label{De0}
\end{equation}


Error $\Delta e$ can be understood in terms of Figure~\ref{fig:phport}. 
This function measures the distance between the green X and the points on the other orbits at $\omega=\pm \pi/2$.
We are interested in how much the orbits in Figure~\ref{fig:phport} converge from "initial" values near the origin  to the values near the green X. As is evident from these plots, there is significant convergence for orbits both inside
and outside the separatrix. As is also seen in Figure~\ref{fig:phport}, the convergence is stronger
with increasing $i_{\rm init}$.

\begin{figure}
\includegraphics[width=8 cm]{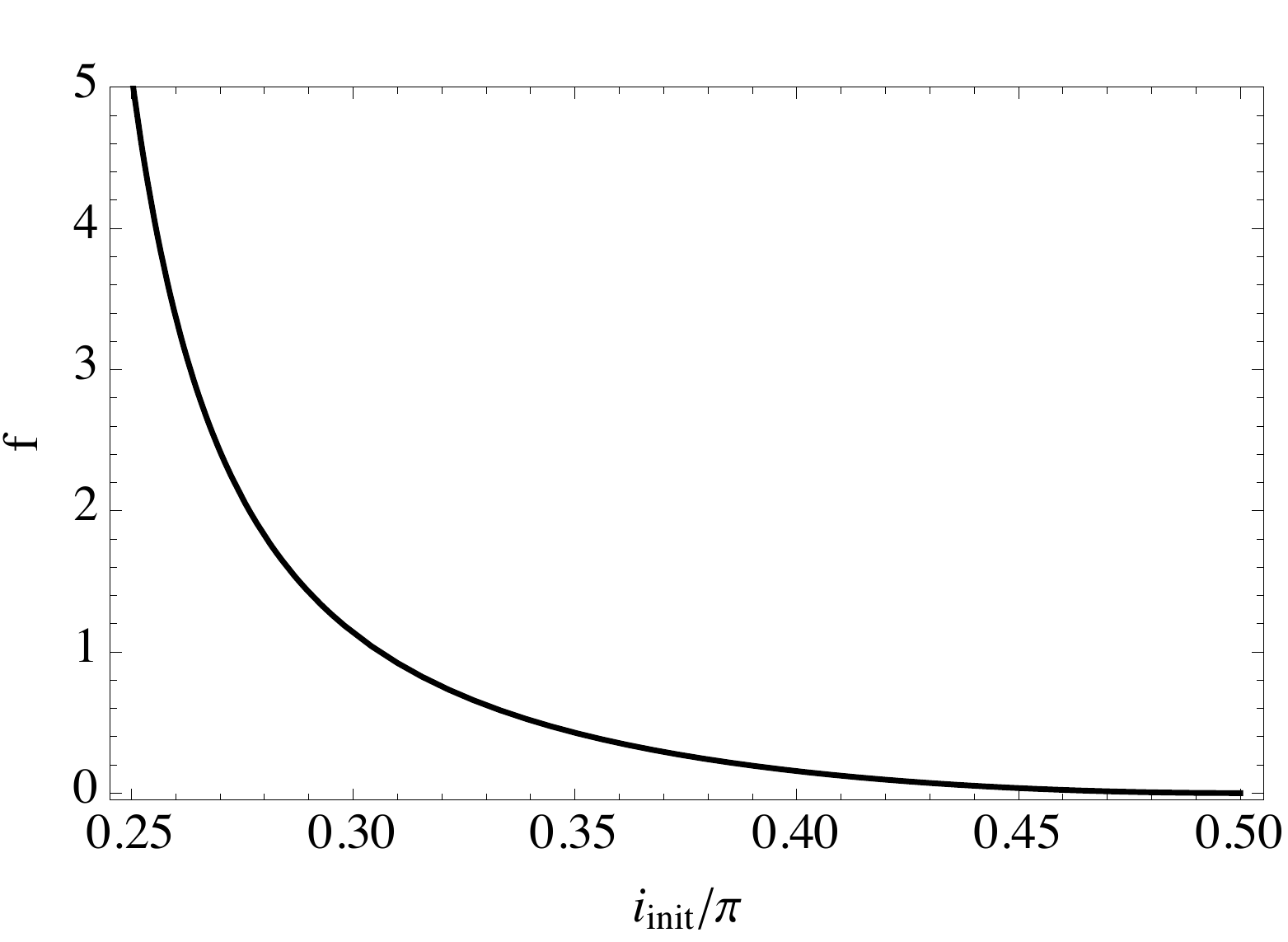}
\caption{Plot of function $f$ that is defined by Equation (\ref{f}).
 }
\label{fig:f}
\end{figure}

We determine $\Delta e (e_{\rm init}, i_{\rm init}, \omega_{\rm init})$ analytically using a series expansion
in $e_{\rm init}$.
For the orbits outside the separatrix with $\omega_{\rm init}=0$ as shown in Figure~\ref{fig:phport}, we have to lowest
order that
\begin{equation}
\Delta e (e_{\rm init}, i_{\rm init}, \omega_{\rm init}=0) = f(i_{\rm init}) \, e_{\rm init}^2 \label{deout}
\end{equation}
where 
\begin{equation}
f( i_{\rm init}) = \frac{ 25 \sin^2{(2  i_{\rm init}})}{2 \sqrt{6} (1-5 \cos{(2  i_{\rm init})})^{3/2}}. \label{f}
\end{equation}
The errors decrease rapidly with decreasing initial eccentricity as $e_{\rm init}^2$.
Function $f( i_{\rm init})$ is plotted in Figure~\ref{fig:f}.
For $i_{\rm init} \simeq \pi/2$, 
we have that
\begin{equation}
f( i_{\rm init}) = \frac{25}{18} \left(\frac{\pi}{2} -  i_{\rm init} \right)^2,
\end{equation}
which shows the error drops rapidly as the initial inclination approaches $\pi/2$.
At the KL critical angle, $i_{\rm crit} = 0.5 \arccos{(0.2)} \simeq 0.217 \pi  \simeq 39.2^\circ$, the denominator
vanishes in Equation (\ref{f}). Just above this critical angle, we have that
\begin{equation}
f( i_{\rm init}) = \frac{1}{ 6^{1/4} 4 ( i_{\rm init}-i_{\rm crit})^{3/2} },
\end{equation}
which shows that the error grows as $i_{\rm init}$ approaches  $i_{\rm crit}$, as seen in Figure~\ref{fig:f}.
In this regime, $\Delta e \ll 1$ for $e_{\rm init} \ll  ( i_{\rm init}-i_{\rm crit})^{3/4}$. 

Orbits inside the separatrix with $\omega_{\rm init} =\pi/2$ all have the same maximum eccentricity, independent of initial eccentricity and inclination and therefore $\Delta e=0$, as seen in Figure~\ref{fig:phport}. 
Orbits that begin inside the separatrix in Figure~\ref{fig:phport} and have $\omega_{\rm init} \ne \pi/2$ cross that separatrix.
Therefore, the heavy black line  is not actually a separatrix for such orbits. For orbits that begin inside the separatrix in Figure~\ref{fig:phport} and have $\omega_{\rm init} \simeq \pi/2$,
we obtain in a  low order series approximation that
\begin{equation}
\Delta e (e_{\rm init}, i_{\rm init}, \omega_{\rm init}) = f(i_{\rm init}) \, \left(\frac{\pi}{2}-\omega_{\rm init} \right)^2 \, e_{\rm init}^2 \label{dein}
\end{equation}
with $f$ defined by Equation (\ref{f}).

\begin{figure}
\includegraphics[width=8 cm]{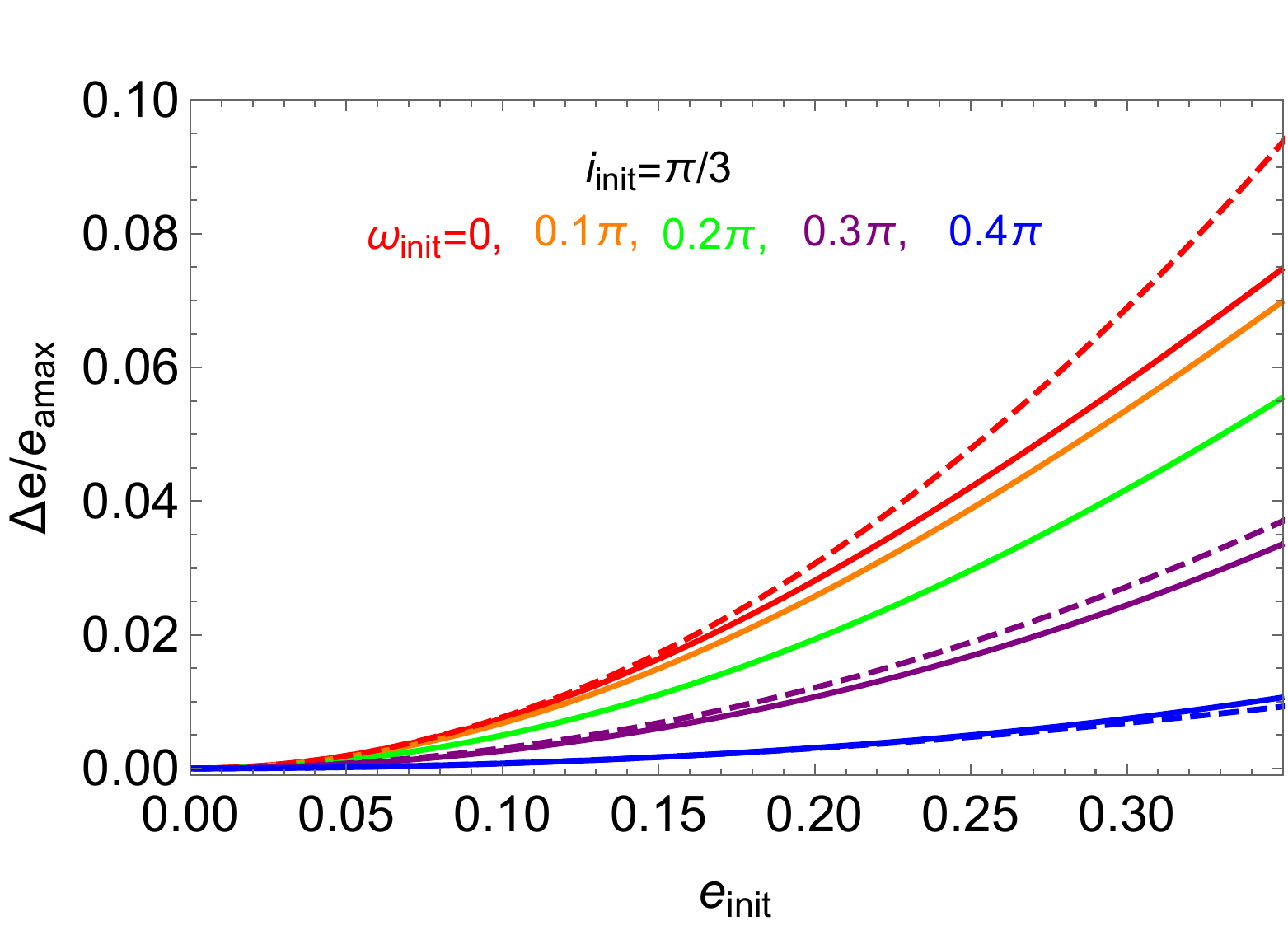}
\caption{Plot of fractional error function $\Delta e/e_{\rm amax}$  defined by Equations (\ref{eamax}) and  (\ref{De0}) 
as a function of $e_{\rm init}$ with $i_{\rm init}=\pi/3$.
The solid lines are determined numerically for the five indicated values of $\omega_{\rm init}$.
The red, orange, and green lines are for orbits that begin outside the separatrix in the middle panel of Figure~\ref{fig:phport}.
The purple and blue lines are for orbits that begin inside that separatrix.
The red dashed line is based on the analytic approximation given by Equation (\ref{deout}). The purple and blue dashed lines are based on the analytic approximation given by Equation (\ref{dein}). 
 }
\label{fig:de0}
\end{figure}

Figure~\ref{fig:de0} plots the fractional error at maximum eccentricity $\Delta e (e_{\rm init},  i_{\rm init}=\pi/3, \omega_{\rm init})/ e_{\rm amax}$ given by Equations (\ref{eamax}) and (\ref{De0}) as a function of $e_{\rm init}$   for five values of $\omega_{\rm init}$.
The solid lines plot the numerically determined $\Delta e/e_{\rm amax}$ values. 
The dashed lines are based
on the analytic approximations provided by Equations (\ref{deout}) and (\ref{dein}) for orbits that begin outside and 
inside the separatrix in the middle panel of Figure~\ref{fig:phport}, respectively.
The analytic approximations agree well with the numerical values of $\Delta e/ e_{\rm amax}$  for small $e_{\rm init}$, as expected.
The  $\Delta e/e_{\rm amax}$ values for orbits that begin inside the separatrix (purple and blue) are smaller than those outside (red, orange, and green).  

In  Figure~\ref{fig:de0} the values of $\Delta e/e_{\rm amax}$ at fixed $e_{\rm init}$ decrease monotonically with increasing $\omega_{\rm init}$ for $0 \le \omega_{\rm init} < \pi/2$.
Also, $\Delta e/e_{\rm amax}$ is symmetric  about $\omega_{\rm init}=0$ and about $\omega_{\rm init}=\pi/2$. That is, 
\begin{eqnarray}
\Delta e(e_{\rm init},  i_{\rm init}, \ \omega_{\rm init})&=& \Delta e(e_{\rm init},  i_{\rm init}, - \omega_{\rm init}),\\
\Delta e(e_{\rm init},  i_{\rm init}, \ \omega_{\rm init})&=& \Delta e(e_{\rm init},  i_{\rm init}, \pi - \omega_{\rm init}).
\end{eqnarray}
Consequently, the results in Figure~\ref{fig:de0} cover values of $\omega_{\rm init}$ in all four quadrants, 
$0 \le \omega_{\rm Init} < 2 \pi$.


\begin{figure}
\includegraphics[width=8 cm]{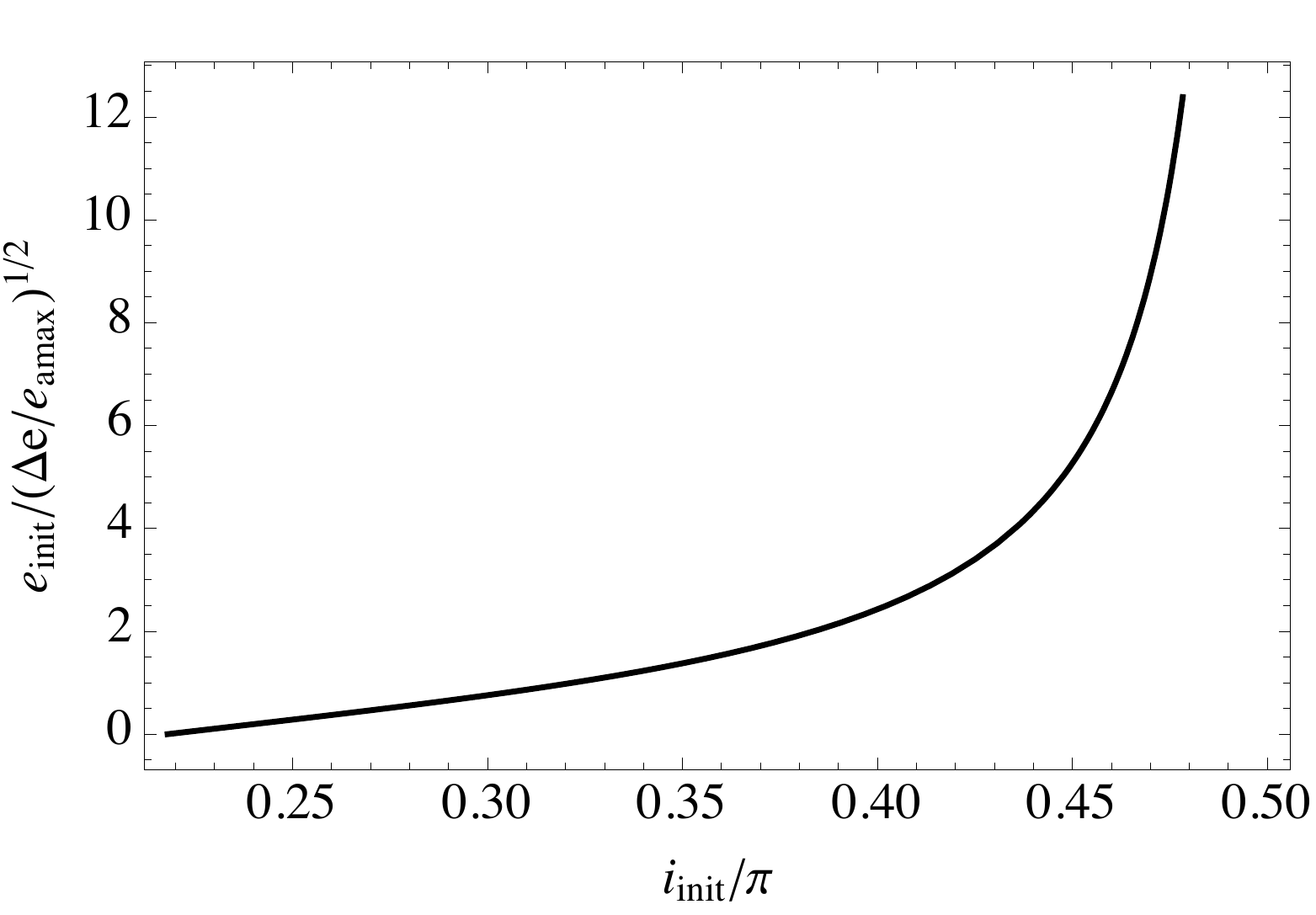}
\caption{Plot of  $e_{\rm init}/\sqrt{\Delta e_{\rm init}/e_{\rm amax} }$ 
as a function of $i_{\rm init}$ that is defined by Equation (\ref{err}).
 }
\label{fig:err}
\end{figure}

As seen in Figure~\ref{fig:de0}, the errors in using the analytic solution with nonzero $e_{\rm init}$ are largest for  the orbits with $\omega_{\rm init}=0$. Combining Equations (\ref{eamax}),  (\ref{deout}), and (\ref{f}), we then estimate
the  $e_{\rm init}$ value  that is small enough to reach a given level of fractional error $\Delta e_{\rm init}/e_{\rm amax}$
as 
\begin{equation}
e_{\rm init} =  \frac{\sqrt{2}(1- 5 \cos{( 2 i_{\rm init} ))}}{5\sin{(2 i_{\rm init})}} \sqrt{\frac{\Delta e}{e_{\rm amax}} }. \label{err}
\end{equation}
Figure~\ref{fig:err} plots the ratio $e_{\rm init}/\sqrt{\Delta e_{\rm init}/e_{\rm amax}}$ as a function of $i_{\rm init}$.
To achieve a small fractional error $\Delta e_{\rm init}/e_{\rm amax} = 0.01$ 
 with an initial tilt of $i_{\rm init} = \pi/4$ requires $e_{\rm init} \la 0.03$,
for $i_{\rm init} = \pi/3$ requires $e_{\rm init} \la 0.1$,
and for $i_{\rm init} = 4 \pi/ 9 = 80 ^\circ$ only requires $e_{\rm init} \la 0.5$.

\section{Summary} 
\label{sec:sum}

This paper considers the case of an initially circular orbit of a test particle around a member of a binary system. The orbit is significantly inclined
with respect to the binary orbital plane. Such a particle undergoes Kozai-Lidov oscillations that
have been the subject of many previous studies.
The companion is assumed
to lie on an orbit that is far outside the orbit of the test particle. 
Under this (quadrupole) approximation,
Equations (\ref{es}), (\ref{is1}),  (\ref{omegas1}), and (\ref{Omegas}) provide an exact analytic
solution to the nonlinear secular evolution of the particle's
orbital elements $e$, $i$, $\omega$, and $\Omega$ given
by Equations (\ref{e}) - (\ref{Omega}). The solution is expressed in terms of simple
trigonometric and hyperbolic functions of time. 
In the case that the particle orbit inclination angle deviates slightly  from being perpendicular to the binary orbital plane by amount $0< \theta \ll 1$, the inclination, argument of periapsis, and longitude of the ascending node undergo large changes on a timescale that is shorter than the eccentricity evolution timescale by about a factor of  $\theta$ (Section \ref{sec:hii}).

The analytic solution extends to all inclinations by taking the real part of $q$ in Equation (\ref{q}) (Section \ref{sec:loi}).
As discussed in Section \ref{sec:num}, the analytic solution agrees well with 
a numerically determined solution. For given initial inclination value $i_{\rm init}$,
the analytic solution determines $\omega_{\rm init}$ given by Equation (\ref{ominit}).
Numerical solutions that begin with other values of $\omega_{\rm init}$
approach the analytic solution for $\omega(t')$  with small values of initial eccentricity $e_{\rm init}$ (see Figure~\ref{fig:omt}).  

In Section \ref{sec:e0} we determine the error in using the maximum eccentricity provided by the analytic solution  for cases
with nonzero initial eccentricity as function of $e_{\rm init}, i_{\rm init}$, and $\omega_{\rm init}$.
The errors are quadratic in $e_{\rm init}$  for small $e_{\rm init}$
 (see Figure~\ref{fig:de0}). The errors increase as $i_{\rm init}$ approaches the KL critical angle and
drop rapidly as $i_{\rm init}$ approaches $\pi/2$ (see Figure~\ref{fig:f}). 
The initial eccentricity required to reach a given fractional error in maximum eccentricity depends
on the initial inclination  (see Figure~\ref{fig:err}).
In the case of an initial inclination of $\pi/3$, an error  of 1\%  at maximum eccentricity
occurs for initial eccentricities of  about 0.1.
The analytic solution in Section \ref{sec:sol} provides good accuracy for a range of initial conditions that
broadens at higher initial inclinations.

\section*{Acknowledgements}
I thank the referee for suggestions that led to improvements in the analysis of convergence to the analytic solution.
I acknowledge support  through NASA XRP grant 80NSSC19K0443. I thank Rebecca Martin and Dan Romik for useful discussions.

\section*{Data availability}
The data underlying this article will be shared on reasonable request to the author.



\bibliographystyle{mnras}
\bibliography{main} 

\appendix
\section{Nonuniform Nodal Precession: The Gudermannian Function and the Mercator Projection }
\label{sec:mercator}

\begin{figure}
\includegraphics[width=8 cm]{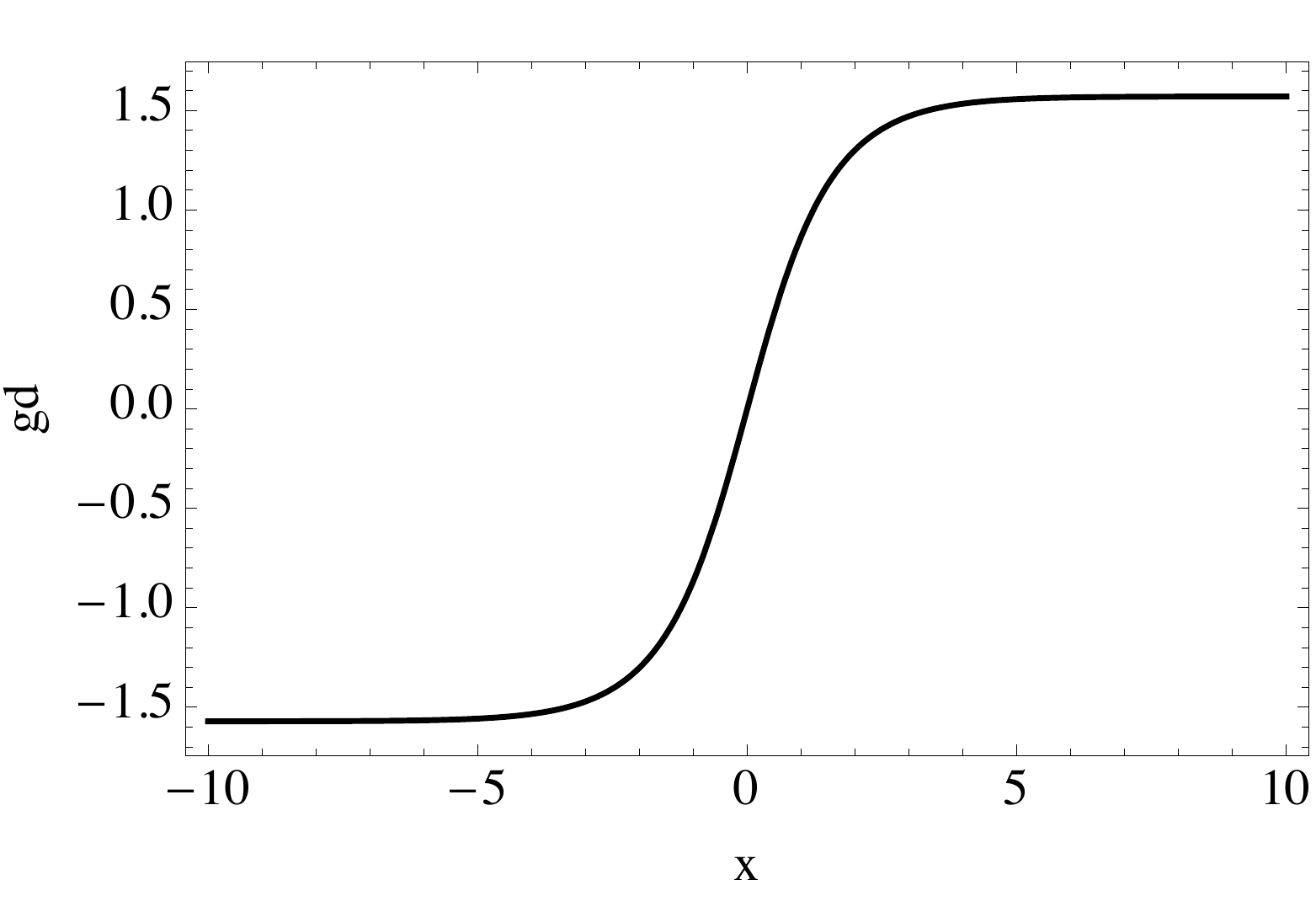}
\caption{Plot of the Gudermannian function $\gd(x)$.
 }
\label{fig:gd}
\end{figure}

We discuss here the nonuniform nodal precession term that is associated with Kozai-Lidov oscillations and appears in Equation (\ref{Omegas})
\begin{equation}
\Omega_{\rm{KL}}(t') = -\arctan{\left(\frac{q \tanh{(q \, t')}}{2 \cos{(i_{\rm init})} } \right)}. \label{OmegaKL}
\end{equation}
The Gudermannian function is defined as
\begin{equation}
\gd(x) = \int_0^x \sech(u)  \, du
\end{equation}
and can be expressed as
\begin{equation}
\gd(x) = 2 \arctan{\left(\tanh{\left(\frac{x}{2} \right)} \right)}. \label{gd}
\end{equation}
The function is plotted in Figure~\ref{fig:gd}.
In general, Equation (\ref{OmegaKL}) cannot be expressed exactly in terms of  Equation (\ref{gd}).
However, it can be related in an approximate way by requiring the function derivatives match at $t'=0$ and
that the function values match at large $|t'|$.
The result is that
\begin{equation}
\Omega_{\rm{KL}}(t')  \simeq c_1 \gd(c_2 \, t' ), \label{OmKLa}
\end{equation}
where
\begin{eqnarray}
c_1 &=& -\frac{2}{\pi} \arctan{ \left( \frac{q}{ 2 \cos{(i_{\rm init})} } \right) },\\
c_2 &=& \frac{\pi q^2}{4 \cos{(i_{\rm init})} \arctan{ \left( \frac{q}{ 2 \cos{(i_{\rm init})} } \right) }}.
\end{eqnarray}
The approximation is fairly accurate for typical parameters in the Kozai-Lidov regime.
In particular, Equation (\ref{OmKLa}) is an exact equality for $q=2 \cos(i_{\rm init})$  in Equation (\ref{OmegaKL}) that occurs for
 $i_{\rm init} = \arccos{(\sqrt{3/7})} \simeq 49.1^\circ$.

The Mercator projection is frequently used for constructing maps of the surface of the Earth. 
The projection transforms points on the surface with longitude and latitude $(\lambda, \phi)$ to an $(x,y)$ Cartesian system (the map).
It is based on a (conformal) mapping that preserves
angles between lines. It does not preserve area, resulting in the familiar problem that the high latitude country Greenland  
appears to be comparable in size to the entire continent of Africa. 
It can be shown that the inverse Mercator projection of latitude that maps $y$ to $\phi$ is a  Gudermannian function of $y$, see \url{https://en.wikipedia.org/wiki/Mercator_projection}.
Therefore, the $\Omega_{\rm{KL}}$ is related to $t'$ by an appropriately scaled Mercator projection 
with  $\Omega_{\rm{KL}}$ playing the role of $\phi$ and $t'$ playing the role of $y$.
The stretched  $y$-extent of Greenland by the Mercator projection (the large value of $dy/d\phi$ at high lattitude $\phi$) 
is analogous to the stretched time $t'$ for larger values of $\Omega_{\rm{KL}}$ (the large value of $dt'/d \Omega_{\rm{KL}}$ at larger $\Omega_{\rm{KL}}$).
This effect is seen in Figure~\ref{fig:gd} as the flatness of the  Gudermannian function (the inverse Mercator projection) at large $|x|$.

\bsp	
\label{lastpage}
\end{document}